\documentclass[aps,prb,twocolumn,superscriptaddress,amsmath,amssymb]{revtex4-2}
\usepackage{graphicx}
\usepackage{dcolumn}% Align table columns on decimal point
\usepackage{bm}% bold math
\usepackage{color}%
\usepackage{hyperref}
\usepackage{wrapfig}
\hypersetup
{
	colorlinks=true,
	linkcolor=blue,
	urlcolor=blue,
	citecolor=blue,
}
\usepackage{mathrsfs}

\usepackage{changes}

\makeatother

\begin{document}

\title{
%Experimental demonstration of strong attenuation within the photonic band gap for multi-connection networks
{Observation of strong attenuation within the photonic band gap of multiconnected networks}
}

\author{Pengbo Zhu}
\author{Runkai Chen}
\author{Xiangbo Yang}
\email{xbyang@scnu.edu.cn}
\author{Yanglong Fan}
\author{Huada Lian}
\affiliation{Guangdong Provincial Key Laboratory of Nanophotonic Functional Materials and Devices, School of Information and Optoelectronic Science and Engineering, South China Normal University, Guangzhou 510006, China}
\author{Zhen-Yu Wang}
\email{zhenyu.wang@m.scnu.edu.cn}
\affiliation{Key Laboratory of Atomic and Subatomic Structure and Quantum Control (Ministry of Education), and School of Physics, South China Normal University, Guangzhou 510006, China}
\affiliation{Guangdong Provincial Key Laboratory of Quantum Engineering and Quantum Materials, and Guangdong-Hong Kong Joint Laboratory of Quantum Matter, South China Normal University, Guangzhou 510006, China}


%\date{\today}

\begin{abstract}
{We theoretically and experimentally study a photonic band gap (PBG) material made of coaxial cables. The coaxial cables are waveguides for the electromagnetic waves and provide paths for direct wave interference within the material. Using multiconnected coaxial cables to form a unit cell, we realize PBGs via (i) direct interference between the waveguides within each cell and (ii) scattering among different cells. We systematically investigate the transmission of EM waves in our PBG materials and discuss the mechanism of band gap formation. We observe experimentally for the first time the wide band gap with strong attenuation caused by direct destructive interference.
}
\end{abstract}


%\keywords{PACS numbers: 42.25.Bs, 11.30.Er, 42.79.Gn}
\maketitle

\section{Introduction}
\label{s1}
Dielectric structures with dielectric constants arranged periodically on the wavelength scale can exhibit photonic band gaps (PBGs), in which the propagation of electromagnetic (EM) waves are inhibited~\cite{Yablonovitch1987Inhibited,John1987strong}. When an incident wave enters a PBGs, it decays rapidly, forming an attenuation mode, a property that is of great importance for the modulation of EM waves~\cite{Askitopoulos2011Bragg,Felbacq2004Bloch,Mario2010Photonic}. Photonic crystals are conventional PBG materials with
PBGs and have attracted large interests both theoretically and experimentally~\cite{Povinelli2001Emulation,Foteinopoulou2012photonic,Lodahl2004controlling,Pradhan1999Impurity,Hach2002Anomalous,Hach2002Long-range,Cheung2004Large,Antonella2003Photonic,Tufarelli2012Dynamics,Ao2009PRB,Fan2013photonic}. However, direct measurement of EM waves in photonic crystals is challenging~\cite{Felbacq2004Bloch,Povinelli2003Toward,Rao2007Single,Antonella2003Photonic}.

A periodic network structure consisting of waveguides has been proposed and localization of states have been observed inside this material~\cite{Zhang1994Wave,Zhang1998Observation}. Such a waveguide network essentially introduces resonant loops, which can be resonated and antiresonated in a certain frequency range~\cite{Zhang1994Wave,Zhang1998Observation,Li2000photonic,C2005Integrated}.  In particular, the waveguide network made of coaxial cables is easy to implement in experiments and allows for a more intuitive study of the EM wave transmission inside the structure, as the amplitude and phase changes of the nodes in each part of the network can be better measured, which will allow for further in-depth study and understanding of the EM wave behavior in the PBGs~\cite{Zhang1994Wave,C2005Integrated,Zhang1998Observation,Li2000photonic}.

It is also convenient to modify the network structure to have interesting features. For example, it was theoretically found that the use of multiple waveguides to connect the same pair of nodes in the network can lead to strong attenuation of the wave propagation within the PBGs~\cite{Wang2007Strong}.  This motivates further theoretical studies in various structures with interesting properties~\cite{Xu2021Ultrawide,Lu2011large,Tang2014super-strong,zheng2019transmission}. Optical waveguide networks can produce rich photon attenuation modes~\cite{Xu2015sufficient}, interesting comb optical transmission spectra~\cite{YANG2013COMB-LIKE,WANG20141200}, extremely wide PBGs~\cite{Wang2007Strong,Cheung2004Large,Li2000photonic}, super-photon localization~\cite{Lu2012large,Xiao2012huge,Xu2015sufficient,Xu2021Ultrawide}. Some of the properties of optical waveguides are applied to microcantilever sensor~\cite{jing2020access}, passive optical device~\cite{Asnawi_2021}, and the physics of ${\mathcal PT}$-symmetry can also be investigated in it~\cite{wu2019reflectionless,zhi2018optical,wang2022quasi,Wu2017}.

%However, some large PBGs have not yet been observed experimentally, while the mechanism for PBG formation rests only on the resonance and antiresonance generated by the scattering geometry~\cite{Zhang1998Observation}.}

In this paper, we experimentally design a one-dimensional (1D) networks system constructed from coaxial cables, where multiple cables are used to realize different paths for the EM waves such that the EM waves can have direct interference within each unit cell [see Fig.~\ref{fig1}]. Changing the relative length ratios of the coaxial cables, we find that direct destructive interference between different paths leads to a very wide PBG. Within this wide band gap, the propagation of EM wave is strongly prohibited. Remarkably, under certain conditions, the incident EM wave in a PBG can not travel through even one unit cell and are almost totally reflected at the boundary of the material, because of the strong attenuation caused by direct interference. 
%\zhu{\wzy{Our results of the new kind of PBG due to direct destructive interference provide a new insight to develop new band gap materials for photons and phonons may provide ideas for other theoretical results.}

The paper is organized as follows: Section II describes the modeling and methodology used to study our PBG material as well as our experimental setup. Section III investigates the energy band structure and attenuation characteristics of our material. We draw our conclusions in Sec. IV.
\section{Model, Theory, and Methods} \label{s2}

\subsection{Network Structure}
\begin{figure}[ht]
	\centering
	\includegraphics[width=\linewidth]{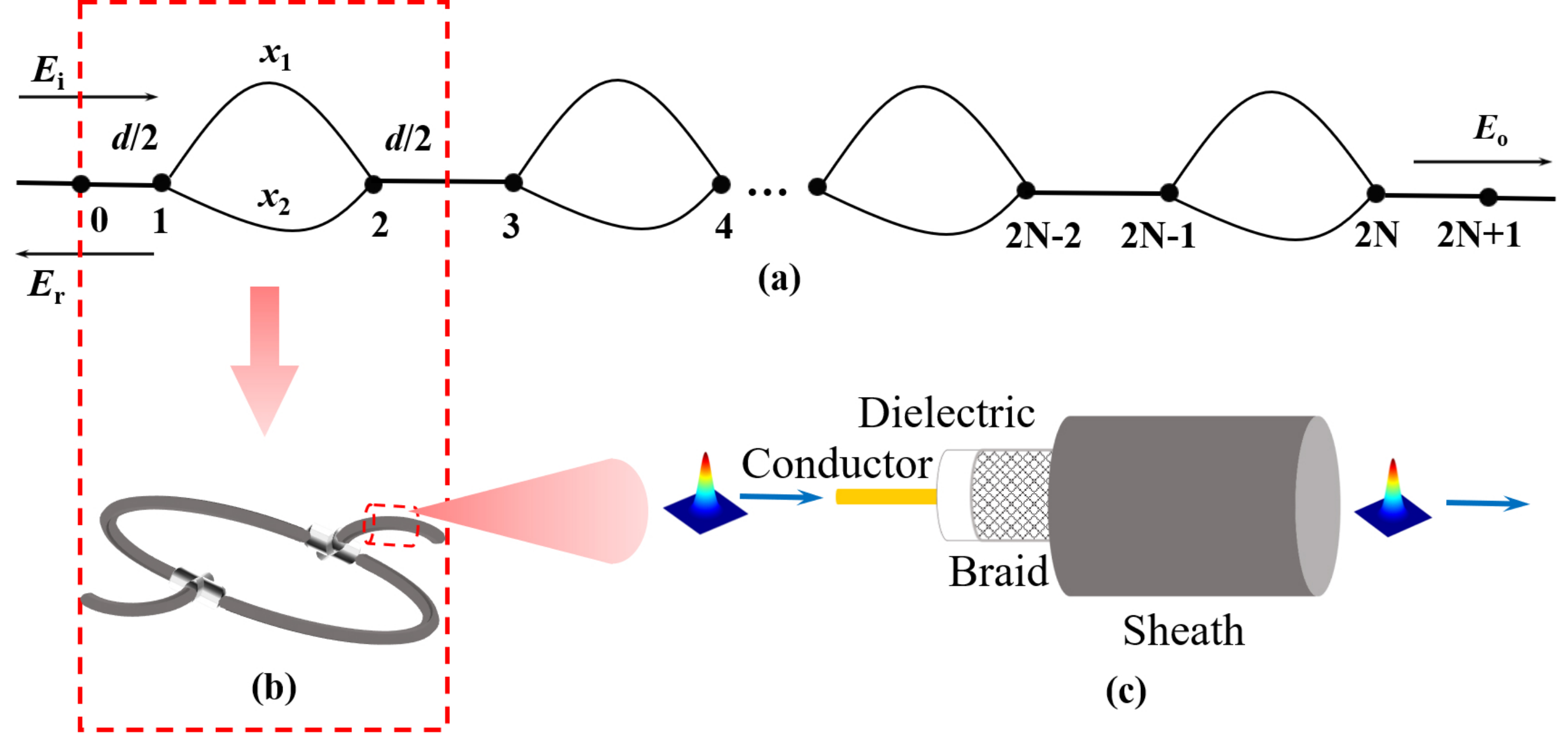}
	\caption{Schematic diagram of the waveguide network model. (a) Schematic structure of a one-dimensional coaxial cable network containing $N$ unit cells. Each unit cell has two waveguides of different lengths ${{x}_{1}}$ and ${{x}_{2}}$ for interference. The structures for interference are linked by coaxial cables of length $d$. An incident EM wave ${{E}_{i}}$ will have a reflected (${{E}_{r}}$) and a transmitted (${{E}_{o}}$) parts. (b) We use connectors to realize the structure of a cell of the model.  (c) The structure of the coaxial cable for our EM waveguides.}
	\label{fig1}
\end{figure}

{Fig.}~\hyperref[fig1]{1(a)} is a schematic structure of our PBG material made of connected waveguides containing {$N$} unit cells. In each cell there are two connected waveguides with lengths ${{x}_{1}}$ and ${{x}_{2}}$ for direct interference, and these interference structures are linked by waveguides of a length $d$. We use coaxial cables to realize the waveguides for EM waves. Fig.~\hyperref[fig1]{1(b)} illustrates  a unit cell in which the cables are connected to each other by connectors. The ports of the connectors can be used to measure the amplitudes of the EM waves. Fig.~\hyperref[fig1]{1(c)} shows the structure of a coaxial cable, which consists of a center conductor, a layer of insulating material, a mesh fabric shield, and an outer sheath. 

Compared with traditional photonic crystals, the structure of our networks is easier to realize; the use of coaxial cables is more flexible from the measurement point of view. In photonic crystals, EM waves are measured mainly on the surface of the material~\cite{Felbacq2004Bloch,Fan2013photonic,Zhang1998Observation}, whereas in the waveguide networks, one can measure the phase and amplitude of EM waves inside any unit cells, thus bringing convenience to experimental studies.

\subsection{Transmission line networks}
\begin{figure}[ht]
	\centering
	\includegraphics[width=0.85\linewidth]{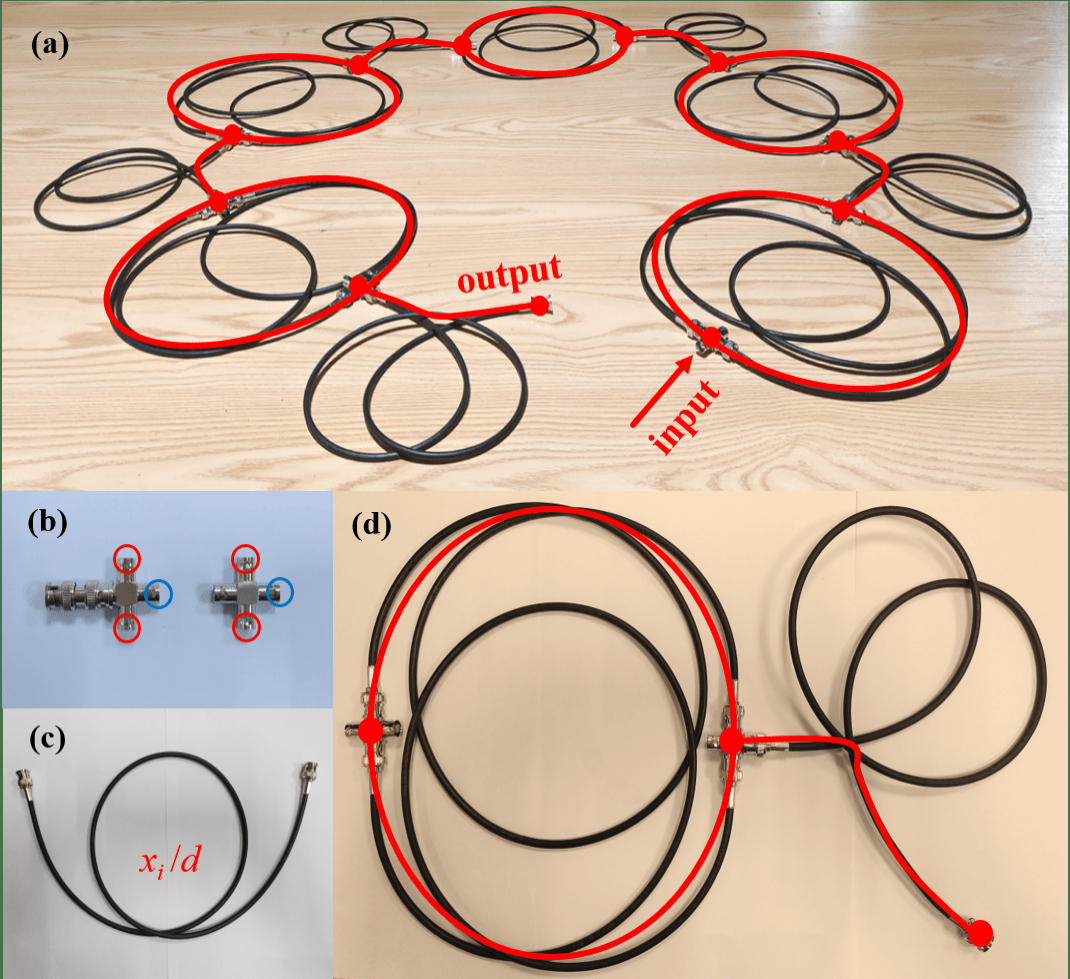}
	\caption{Transmission line networks. (a) Photograph of an experimental sample, which has a size of {$N=5$}. The nodes are marked with red dots and the red curves indicate the connectivity of the network. (b) Nodes formed by connectors.Red circles mark the ports connecting the upper and lower arms of the interference structure. Blue circles indicate the ports for connecting the unit cells or for connecting measurement cables. (c) A coaxial cable waveguide with connectors at the ends. (d) A unit cell consisting of cables and connectors.}
	\label{fig2}
\end{figure}

Pictures of our experimental samples are shown in Fig.~\ref{fig2}, with  Fig.~\hyperref[fig2]{2(a)} showing a network system consisting of 5 unit cells. The connectors used for our experiments are shown in Fig.~\hyperref[fig2]{2(b)}. The red ports of the connector are used to connect the upper and lower arms of the interference structure in each unit cell, while the blue ports are used to connect the interference structures of successive unit cells or for measurement of the EM waves. In our experiments, we used a coaxial cable (type RG58C/U) with the waveguide length shown in Fig.~\hyperref[fig2]{2(c)}. The details of the cable-node connection can be seen in the single-cycle unit cell formed by the connectors and the cables, as shown in Fig.~\hyperref[fig2]{2(d)}.

\subsection{Equations for transmission line networks with single materials}
For our networks system shown in Fig.~\ref{fig1},the propagation of EM waves in a coaxial cable line satisfies the following homogeneous wave equation~\cite{Zhang1998Observation,Li2000photonic}:
\begin{align}\label{e1}%eq1
 {\frac{{{\partial}^{2}}{{\psi }_{nm}}(x)}{\partial {{x}^{2}}}=\frac{\varepsilon {{\omega }^{\text{2}}}}{{{c}_{0}}^{2}}{{\psi }_{nm}}(x),}
\end{align}
where ${{\psi}_{nm}}(x)$ denotes the voltage wave function of any segment between nodes $n$ and $m$, $x$ is the distance measured from node $m$, $\omega =2\pi f$ is the angular frequency of EM wave, ${c}_{0}$ is the wave speed of EM wave in vacuum, $\varepsilon ={\varepsilon }'+i{\varepsilon }''$ is the relative permittivity of the dielectric for the coaxial cables, and $\varepsilon'$ and $\varepsilon''$ are the real and imaginary parts of the relative permittivity, respectively. The EM wave function can be written as a linear combination of two plane waves traveling in opposite directions~\cite{Li2000photonic,Wang2007Strong,hu2017super,Lu2011large,Lu2012large}:
\begin{align}\label{e2}%eq1
{{\psi }_{nm}}={{\alpha }_{nm}}{{e}^{i\kappa x}}+{{\beta }_{nm}}{{e}^{-i\kappa x}},
\end{align}
where $i=\sqrt{-1}$ is the imaginary unit, $\kappa =\frac{\omega }{c}\sqrt{\varepsilon }=\frac{\omega }{c}\sqrt{\varepsilon '+i\varepsilon ''}$~\cite{Zhang1998Observation,Li2000photonic}. In the case of $\varepsilon ''\ll \varepsilon '$, it is deduced that $\kappa $ has the simple form that $\kappa \cong k+(\text{1}/2L)i$, where $k={\omega \sqrt{\varepsilon ''}}/{{{c}_{0}}}\;$ and $L={\varepsilon '}/{k\varepsilon ''}\;$ is the absorption length. For any node $n$ of the network, its wave function is continuous at the node~\cite{Li2000photonic,Wang2007Strong,Xiao2012huge,Lu2011large,Lu2012large}:
\begin{align}\label{e3}%eq10
   \begin{array}{*{35}{l}}
   {{\left. {{\psi }_{nm}} \right|}_{x=0}}={{\psi }_{n}}, \\ 
  {{\left. {{\psi }_{nm}} \right|}_{x={{x}_{nm}}}}={{\psi }_{m}}, 
\end{array}
\end{align}
where ${{\psi }_{n}}$ and ${{\psi }_{m}}$ represent the wavefunction at the nodes $n$ and $m$, respectively. $x_{nm}$ is the length of a waveguide connecting the nodes $n$ and $m$. With the use of Eqs.~\eqref{e2} and \eqref{e3}, one obtains~\cite{Lu2011large,Lu2012large,Xiao2012huge}:
\begin{align}\label{e4}%eq1
{{\psi }_{nm}}={{\psi }_{n}}\frac{\sinh \left[ i\kappa ({{x}_{nm}}-x) \right]}{\sinh (i\kappa {{x}_{nm}})}+{{\psi }_{m}}\frac{\sinh (i\kappa x)}{\sinh (i\kappa {{x}_{nm}})},
\end{align}
At each node $m$, the wave function is continuous and the derivative of the wave function at node $m$ gives the flux conservation condition~\cite{Wang2007Strong,Lu2011large,Xiao2012huge}:
\begin{align}\label{e5}%eq1
{{\left. \sum\limits_{n}{\frac{\partial {{\psi }_{nm}}(x)}{\partial x}} \right|}_{x=0}}=0,
\end{align}
where $n$ is the sum of all nodes directly connected to $m$. Substituting Eq.~\eqref{e4} into Eq.~\eqref{e5}, one gets~\cite{Zhang1994Wave,Zhang1998Observation,Li2000photonic,hu2017super,Lu2011large,Wang2007Strong,Lu2012large}:
\begin{align}\label{e6}%eq1
-{{\psi}_{n}}\sum\limits_{m}{\coth(i\kappa{{x}_{nm}})}+\sum\limits_{m}{\frac{{{\psi }_{m}}}{\sinh (i\kappa {{x}_{nm}})}}=0.
\end{align}
for the network consisting of coaxial cables without considering dissipation. The coth and sinh denote hyperbolic tangent and hyperbolic sine functions, respectively. We have added a constant $\gamma$ to the $i\kappa {{x}_{nm}}$ term of Eq.~\eqref{e6} to account for the additional losses caused by the connectors if we consider the dissipation of the material.

We obtain the absorption length $L$ of the cables and the attenuation $d$ of the connectors by using the calculated transmission coefficients which fit the measured transmission coefficients in a single cell. Using the cable specifications given by the vendor, as well as our experimental measurements and data fitting, we obtained $L/{\rm{m}}=665{{(f/{\rm{MHz}})}^{-0.52}}$ and $\gamma =-0.005$. We used $\varepsilon'=2.3$.

	Therefore, the transmission coefficient $t$ theoretically is calculated using the generalized eigenfunction method to solve the equation, where $t$ is the transmission amplitude of the outgoing wave. The transmittance is ${{\left| t \right|}^{2}}$~\cite{Lu2011large,Lu2012large,Xiao2012huge,hu2017super,Li2000photonic}.

\begin{figure*}
\includegraphics[width=0.75\linewidth]{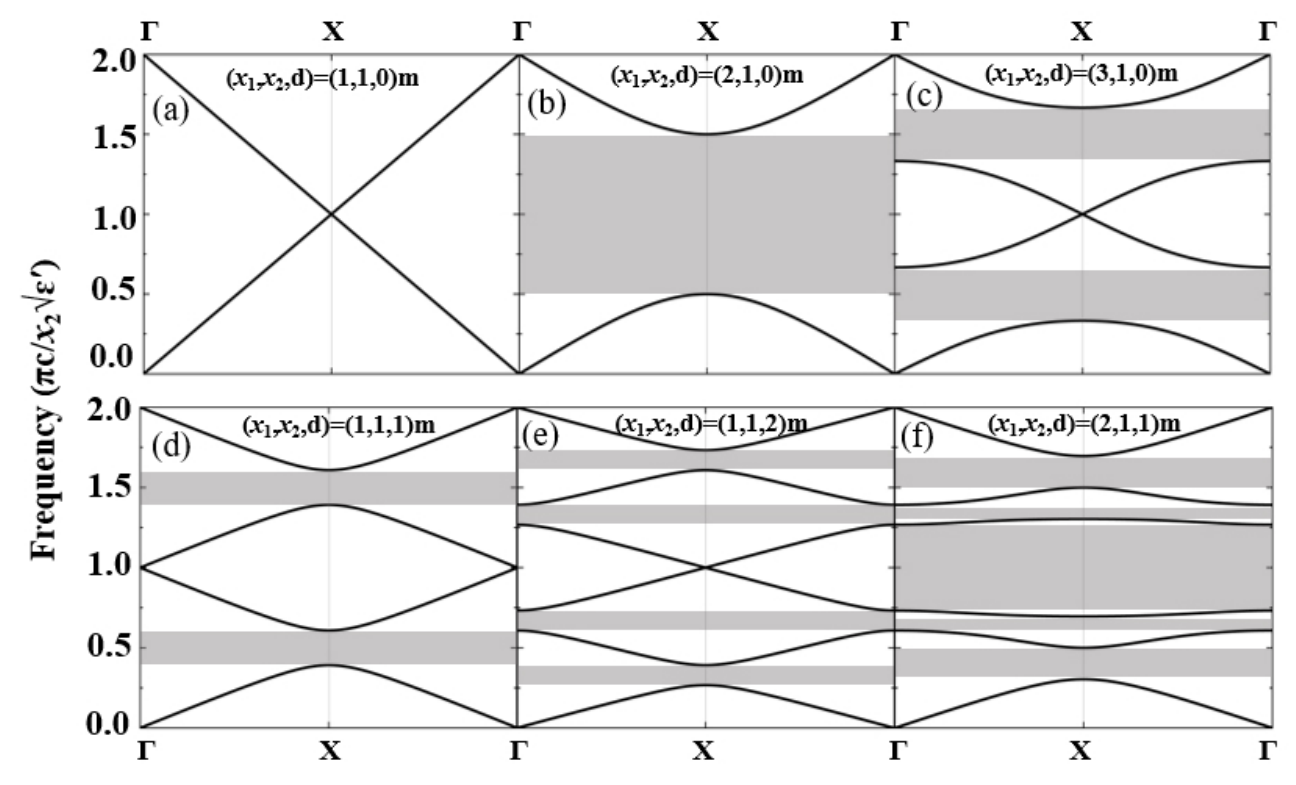}
\caption{\label{fig:wide}Band structures for different lengths of $x_1$, $x_2$, and $d$ , with the gray area indicating the band gap. Here we assume no absorption and losses,i.e., $L=\infty $ and $\gamma =\text{0}$.}
\label{fig3}
\end{figure*}
\subsection{Generalized Floquet-Bloch 's Theorem}
The network equation describes the connection relationship between adjacent nodes in a network. In order to quantitatively analyze the energy bands of the system, we derive the dispersion relation of the network structure. To tackle the problem that in our network system the length of a unit cell is not well defined when the same pair of nodes in the unit cell are connected with cables of different lengths, we use the generalized version of the Floquet-Bloch theorem with the following Bloch function~\cite{Wang2007Strong,hu2017super,Lu2011large,Lu2012large,Xiao2012huge}:
\begin{align}\label{e7}
{\psi }_{n+T}({\Phi})=\psi_{n}({\Phi})e^{i {\Phi} T}.
\end{align}
where the integer $n$ denotes an arbitrarily  node index, the integer $T$ is the period of the network configuration, and $\Phi$ is a dimensionless phase factor. For the 0, 2, and 4 nodes in Fig.~\ref{fig1}, it follows from Eq.~\eqref{e6}:
\begin{widetext}
\begin{equation}
\left\{
\begin{aligned}\label{e8}
\begin{matrix}
   -{{\psi }_{1}}\left[ \coth (i\kappa {{x}_{1}})+\coth (i\kappa {{x}_{2}})+\coth (i\kappa d) \right]+ [\frac{{{\psi }_{0}}}{\sinh (i\kappa d)}+\frac{{{\psi }_{2}}}{\sinh (i\kappa {{x}_{1}})}+\frac{{{\psi }_{2}}}{\sinh (i\kappa {{x}_{2}})} ]=0,  \\
   
   -{{\psi }_{2}}\left[ \coth (i\kappa {{x}_{1}})+\coth (i\kappa {{x}_{2}})+\coth (i\kappa d) \right]+[ \frac{{{\psi }_{3}}}{\sinh (i\kappa d)}+\frac{{{\psi }_{1}}}{\sinh (i\kappa {{x}_{1}})}+\frac{{{\psi }_{1}}}{\sinh (i\kappa {{x}_{2}})} ]=0,  \\
   
   -{{\psi }_{3}}\left[ \coth (i\kappa {{x}_{1}})+\coth (i\kappa {{x}_{2}})+\coth (i\kappa d) \right]+[ \frac{{{\psi }_{2}}}{\sinh (i\kappa d)}+\frac{{{\psi }_{4}}}{\sinh (i\kappa {{x}_{1}})}+\frac{{{\psi }_{4}}}{\sinh (i\kappa {{x}_{2}})} ]=0.  \\
\end{matrix} 
\end{aligned}
\right.
\end{equation}
\end{widetext}
From Eq.~\eqref{e7}, {the wave functions at node 0 and node 4 are related to the wave function at node 2 as}~\cite{Wang2007Strong,Xu2021Ultrawide,Xiao2012huge}:
\begin{align}\label{e9}%eq10
   \begin{array}{*{35}{l}}
   {{\psi }_{0}}={{\psi }_{2}}{{e}^{-i{\Phi}}},  \\
   {{\psi }_{4}}={{\psi }_{2}}{{e}^{i{\Phi}}},  \\
\end{array}
\end{align}
Substituting Eq.~\eqref{e9} into {Eq.~\eqref{e8} gives
\begin{align}\label{e10}%eq3
   \cos \Phi=f({{{\omega\sqrt{\varepsilon'}}/{c}}_{0}})\equiv\frac{{{\text{A}}^{2}}{{\text{B}}^{2}}-C-2B\sinh (i\kappa d)}{2B\left[ \sinh (i\kappa {{x}_{1})}+\sinh (i\kappa {{x}_{2}}) \right]},
\end{align}
}where 
\begin{align}
   \begin{array}{*{35}{l}}
  A =\coth (izd)+\sum\limits_{i=1}^{2}{\coth (iz{{x}_{i}})}, \\
B =\sinh (izd)\prod\limits_{i=1}^{2}{\sinh (iz{{x}_{i}})}, \\
C =  {{\sinh }^{2}}(iz{{x}_{1}}){{\sinh }^{2}}(iz{{x}_{2}}) \nonumber\\
  +{{\sinh }^{2}}(iz{{x}_{1}}){{\sinh }^{2}}(izd)+{{\sinh }^{2}}(iz{{x}_{2}}){{\sinh }^{2}}(izd).
\end{array}
\end{align}
%\begin{align}\label{e13}%eq3

%\end{align}

From Eq.~\eqref{e10} we obtain the band structure of our networks. Because of Eq.~\eqref{e7}, when the solution
$\Phi$ is a complex number with a non-zero imaginary part $\rm{Im}(\Phi)$, the propagation of EM wave among the network nodes has a decay rate given by the amplitude of $\rm{Im}(\Phi)$. That is, we have a band gap when $\Phi$ has no real solution in Eq.~\eqref{e10}. When the amplitude of $\rm{Im}(\Phi)$ is larger, the propagation of EM wave among the network nodes has a stronger attenuation. On the other hand, a real $\Phi$ solution corresponds to a propagation mode among the network nodes. Therefore, we obtain the dispersion relation for the energy bands 
%(that is, the relation between $\Phi$ and the energy of the photon, which is $\hbar\omega$ with $\hbar$ being the Plank's constant) 
from the existence of solutions with a real $\Phi$~\cite{Wang2007Strong}.

%the dispersion relation for this network according to the existence condition for nontrivial solutions,\cite{r16}
%\begin{align}\label{e12}%eq3
%   \cos K=f(z)=f({{{\omega\sqrt{\varepsilon'}}/{c}}_{0}}).
%\end{align}

\begin{figure}[ht]
	\centering
	\includegraphics[width=0.95\linewidth]{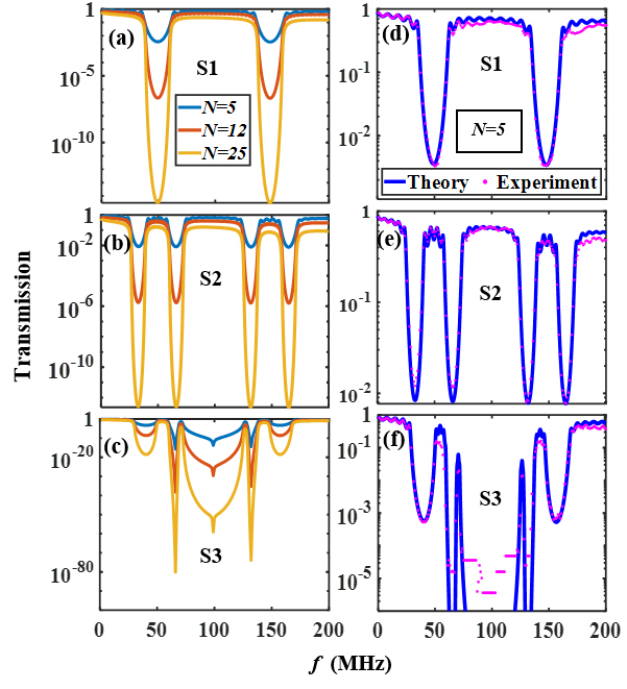}
	\caption{The experimentally measured transmittance $|t|^2$, as well as the theoretically calculated values for the networks S1, S2 and S3 are shown in Fig.~\ref{fig4}. 
The results in Fig.~\ref{fig4} show that the transmittance at the PBGs is getting smaller for a larger size $N$ of the network. It is more interesting to see that the attenuation of 
EM waves within the PBGs is significantly larger for the network S3, which has the ratio $x_1/x_2 =2$, when comparing with other networks S1 and S2 which have $x_1/x_2 =1$. This shows that the destructive interference between the two arms $x_1$ and $x_2$ significantly enhances the formation of PBGs. The results are consistent with the band structures in Fig.~\ref{fig3}}
	\label{fig4}
\end{figure}
\section{Results and discussion}
\label{s3}
\subsection{Observation of large PBGs}
The structure of our network allows us to tune the interference effect between the two arms by changing the lengths  $x_1$ and $x_2$ (see Fig.~\ref{fig1}). Without lost of generality, we assume $x_1\geq x_2$. Furthermore, 
changing the length $d$ of the cables that connect the interference structures together, we can also modify the scattering among different cells. From Eq.~\eqref{e10} and Fig.~\ref{fig3}, we can see that the band structure changes with the ratio between $x_1$ and $x_2$ as well as the length $d$.
The forbidden bands are indicated by gray areas in  Fig.~\ref{fig3}. As demonstrated in Fig.~\ref{fig3}, the ratio between the lengths  $x_1$ and $x_2$ of the two arms has a large influence on the PBGs~\cite{Wang2007Strong,Xu2021Ultrawide,Lu2011large}.

For the case of $x_1=x_2$ there is no destructive interference, and therefore when $d=0$ there is no PBG. When $d>0$ there are small PBGs even $x_1=x_2$, which is due to scattering among different cells. When $x_1/x_2 =2$, there is a strong  destructive interference, and as a consequence, a large band gap is formed.

\subsection{Experimental demonstration of transmittance}

To experimentally probe the system, we connected an Agilent E4438C vector signal generator at one side of the network to input voltage waves. An oscilloscope was connected at the other side of the network  to measure the transmission coefficients and intensities.
Based on the theoretical results shown in Fig.~\ref{fig1}, we prepared different samples of the networks with three different sets of parameters, namely, Structure 1 (S1) with $({{x}_{1}},{{x}_{2}},d)=(1,1,1)$ m, S2 with $({{x}_{1}},{{x}_{2}},d)=(1,1,2)$ m, and S3 with $({{x}_{1}},{{x}_{2}},d)=(2,1,1)$ m.

The experimentally measured transmittance $|t|^2$, as well as the theoretically calculated values for the networks S1, S2 and S3 are shown in Fig.~\ref{fig4}. 
The results in Fig.~\ref{fig4} show that the transmittance at the PBGs is getting smaller for a larger size $N$ of the network. It is more interesting to see that the attenuation of 
EM waves within the PBGs is significantly larger for the network S3, which has the ratio $x_1/x_2 =2$, when comparing with other networks S1 and S2 which have $x_1/x_2 =1$. This shows that the destructive interference between the two arms $x_1$ and $x_2$ significantly enhances the formation of PBGs. The results are consistent with the band structures in Fig.~\ref{fig3}.

\begin{figure}[ht]
	\centering
	\includegraphics[width=\linewidth]{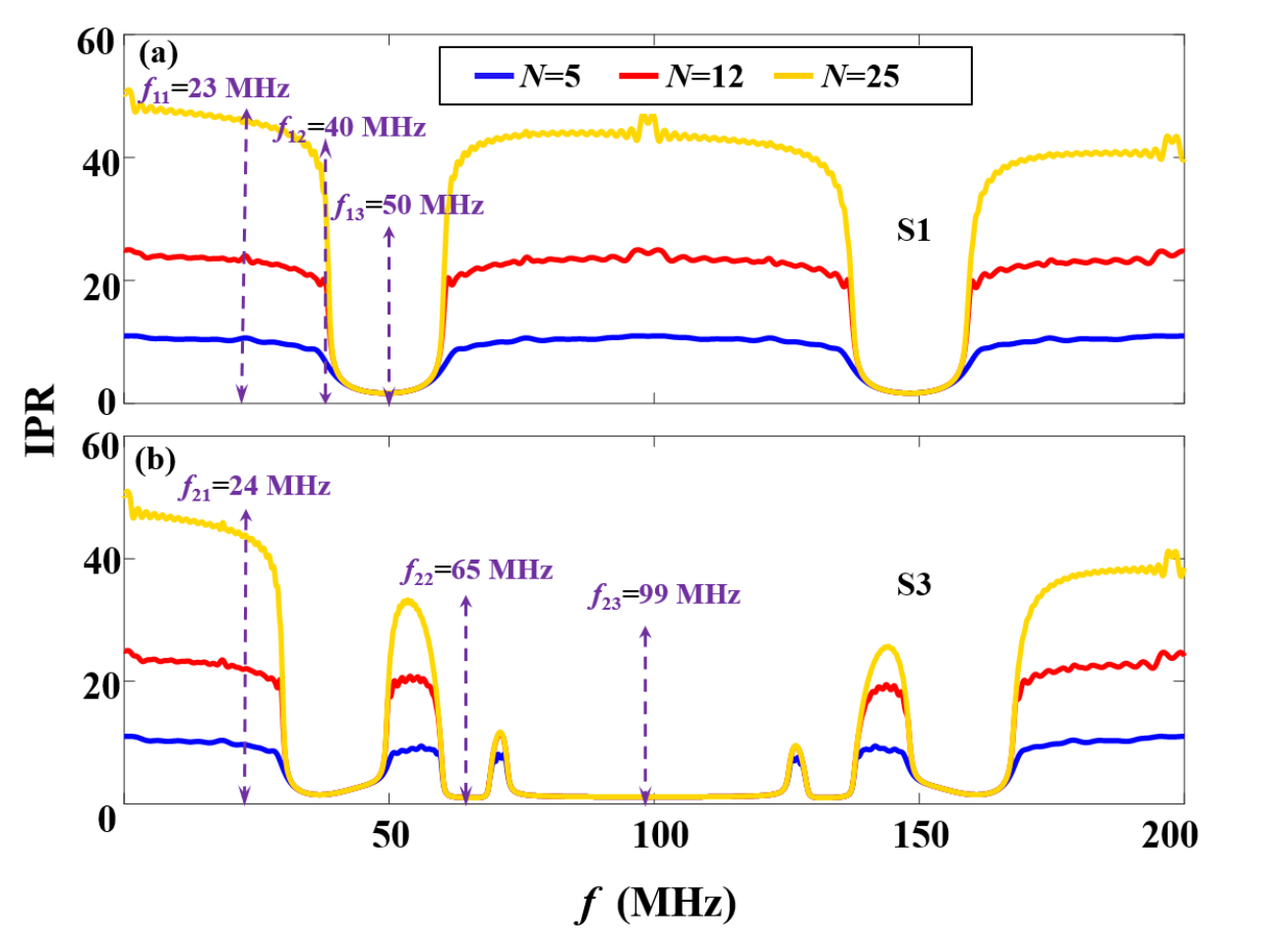}
	\caption{{(a) IPR spectra for the S1 networks with different values of $N$. (b) as (a) but for the S3 networks. }}
	\label{fig5}
\end{figure}
\subsection{Wave Intensity Distribution at Characteristic Frequencies}
In order to investigate the propagation modes of EM waves, we have systematically probed the S1 and S3 networks with various periods $N=5$, 12, and 25. Firstly, we use the inverse participation ratio~\cite{Edwards1972numerical,Zhang1998Observation,hu2017super}:
\begin{equation}
{\text{IPR}={{(\sum\nolimits_{n}{\left| {{\psi }_{n}}^{2} \right|})}^{2}}}/{\sum\nolimits_{n}{{{\left| {{\psi }_{n}} \right|}^{4}}}}\,
\end{equation}
to investigate the wave intensity distribution. For the propagation modes, the value of IPR increases with the size (i.e., $N$) of the network~\cite{Zhang1998Observation}. While for the case that the frequency of a EM wave lying within a PBG, the value of IPR hardly changes with the network size. 

We plot the IPR spectral distributions in Fig.~\ref{fig5}, where we can see that for frequencies near the band edge, IPR increases with the size of the networks. As the frequency enters the gap, the IPR becomes smaller and the size dependence becomes smaller. 
\begin{figure}[ht]
	\centering
	\includegraphics[width=\linewidth]{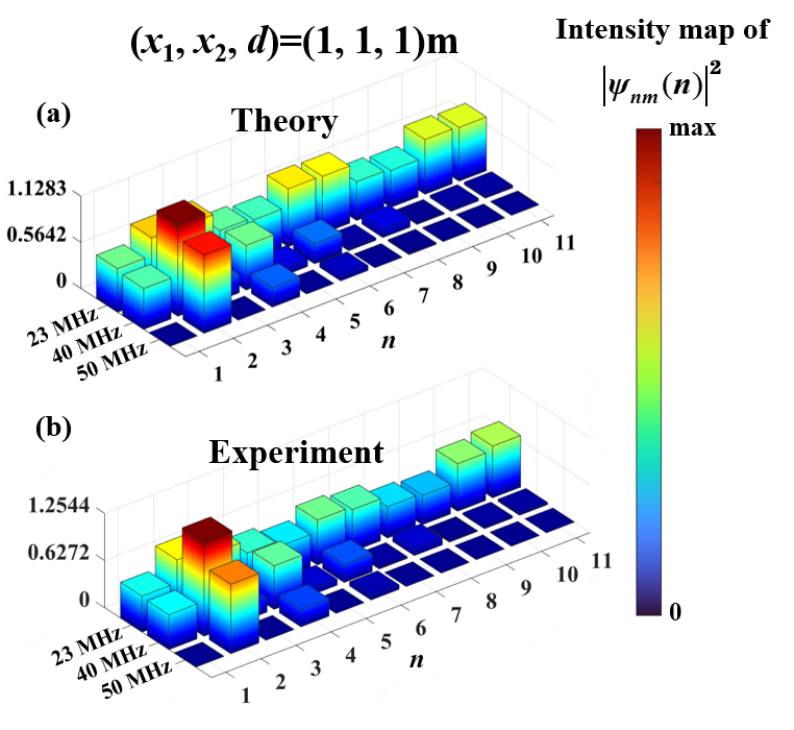}
	\caption{{Intensity of the wave distribution for $N=5$ for different frequencies of incident EM waves in an S1 network. (a) is the simulated results while (b) is the results of experiment.}}
	\label{fig6}
\end{figure}
\begin{figure}[ht]
	\centering
	\includegraphics[width=1\linewidth]{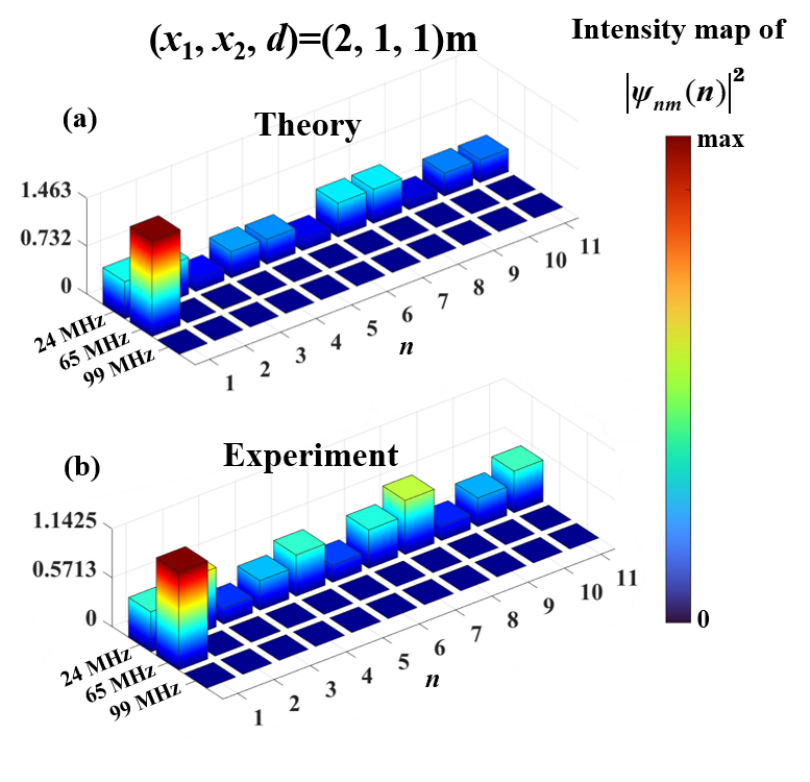}
	\caption{{Intensity of the wave distribution for $N=5$ for different frequencies of incident EM waves in an S3 network. (a) is the simulated results while (b) is the results of experiment.}}
	\label{fig7}
\end{figure}
To have a clear picture of the EM wave distribution inside the network, we further conducted experiments to measure the intensities of the EM waves of all the nodes in the networks, for the EM wave frequencies indicated in Fig.~\ref{fig5}. We show in Figs.~\ref{fig6} and \ref{fig7} the voltage intensities of all the nodes for the networks S1 (with $x_1=x_2$) and S3 (with $x_1=2 x_2$), respectively. One can see that the theoretical results agree with the experimental values.
The results show that at the frequencies when the IPR increases with the network size $N$, the energy of EM wave have a broad distribution, which indicates an extended state. At the frequencies when the IPR is not sensitive to the network size, the intensities are negligible over a broad range of nodes, which suggests an evanescent mode~\cite{Xu2015sufficient}.

\begin{figure}[ht]
	\centering
	\includegraphics[width=\linewidth]{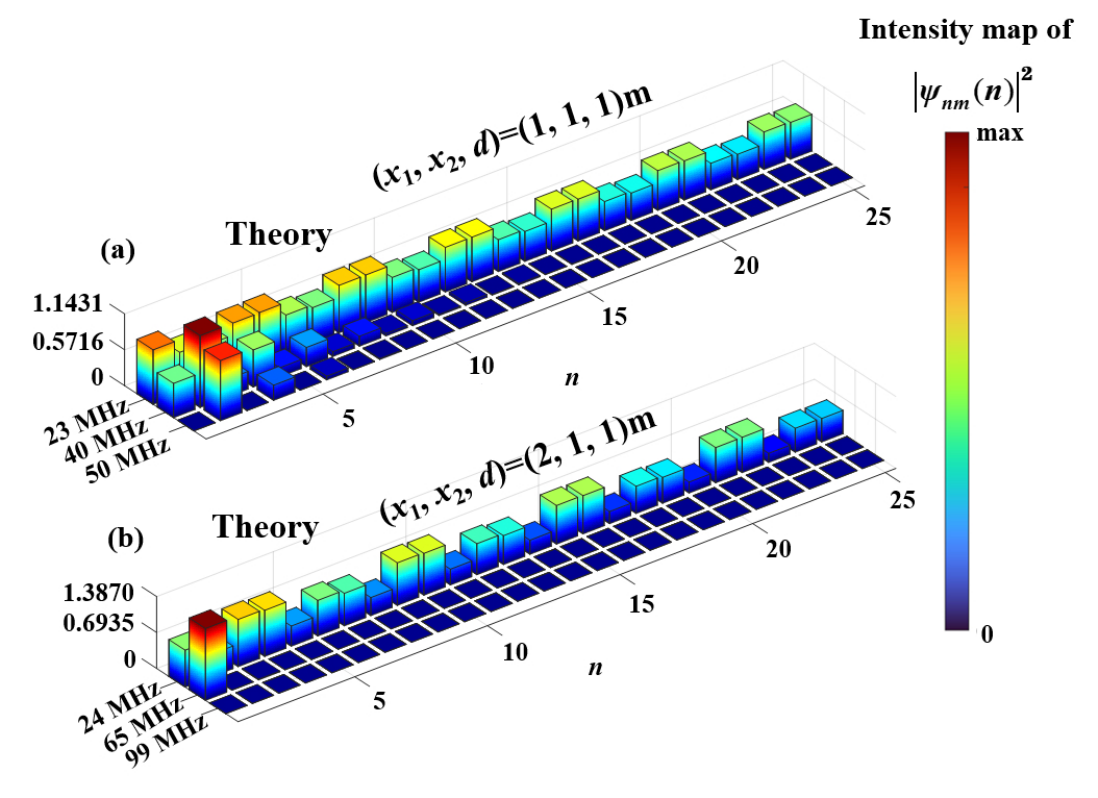}
	\caption{{Simulated wave intensity distribution with  N=12 for different frequencies of incident EM waves. (a)} Intensity distribution map for the S1 network; (b) Intensity distribution map for the S3 network.}
	\label{fig8}
\end{figure}

Note that as we can see in Fig.~\ref{fig7} for the network S3, which has $x_1=2 x_2$,
an ultra strong attenuation of EM wave within a PBG is observed. In particular, for the frequency $f_{23}=99$ MHz close to the center of the PBG, all the nodes within the network have negligible intensities, because the amplitude of the attenuation mode attenuates to zero even within the first unit cell and all the EM wave energy is reflected at the input side. This ultra strong attenuation is due to the direct destructive interference between the two arms of lengths $x_1$ and $x_2$. For the network S1, which has $x_1=x_2$, the attenuation of EM wave within PBGs is weaker, when one compares the results in Figs.~\ref{fig6} and \ref{fig7} as well as the plots in Fig.~\ref{fig8}.
The results show that the PBG due to direct interference between the waveguides within each cell is quite different from the conventional PBGs due to scattering among different cells and provides a much stronger attenuation of EM waves.
\begin{figure}[ht]
	\centering
	\includegraphics[width=1\linewidth]{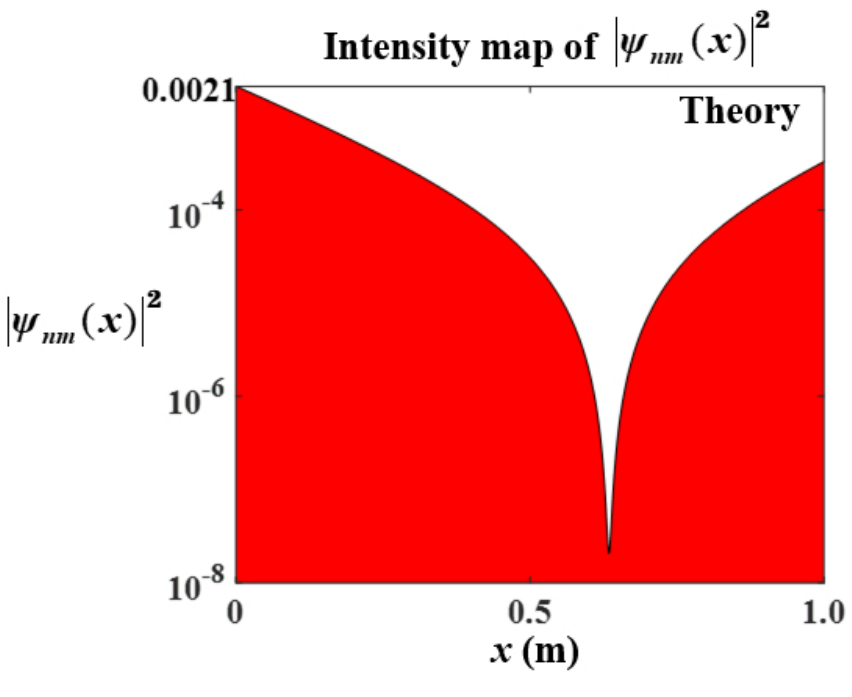}
	\caption{Calculated wave intensity distribution as a function of the location $x$ relative to node $1$ inside a S3 network with $N=5$ for a wave frequency $f_{23}=99$~MHz. The length between the input node 0 and node 1 is 1 m.
	}
	\label{fig9}
\end{figure} 

To further investigate the strong attenuation mechanism of the network of S3 at the frequency $f_{23}=99$ MHz, in Fig.~\ref{fig9} we calculated the wave intensity distribution between the signal generator output and the first contact of the network using the theory of Eq.~\eqref{e4}. We find that the wave intensities in the coaxial cable from the signal output to the first node of the network are all less than 0.0021. Meanwhile, we performed experimental verification and found that the voltage waves are indeed attenuated rapidly in the first section of the coaxial cable. It is clear that this is not due to the geometric scattering structure in the network, but is the result of the interference between the incident and reflected waves in a single coaxial cable.

\section{Conclusions}
\label{s4}
We have theoretically and experimentally study a PBG material made of multiconnected coaxial cables. We have observed for the first time large PBGs  which is due to direct wave interference within each unit cell. From the measured transmission spectra and intensities of EM waves inside the networks, we found that direct interference between two waveguides within the same unit cell provides a new mechanism of PBG formation. In particular, when the length ratio between the two arms $x_1/x_2 =2 $, a PBG with ultra strong attenuation of EM waves is formed due to direct destructive interference. The strong attenuation within the PBG induces an ultra strong reflection of EM waves at the boundary of the network.  Our results demonstrate a new way to realize PBGs and may have applications to the control of various waves (e.g., EM waves or acoustic waves). It would also be interesting to study how the multiconnected waveguides would influence the topological transport in transmission line network~\cite{Jiang2019experimental}.

\begin{acknowledgments}
This work was supported by the National Natural Science Foundation of China (Grant Numbers 11674107, 61475049, 11775083, 61774062, 61771205, {12074131)} and the Natural Science Foundation of Guangdong Province, China {(Grant No. 2021A1515012030)}.
\end{acknowledgments}

\end{document}